# On-device Computation of Single-lead ECG Parameters for Real-time Remote Cardiac Health Assessment: A Real-world Validation Study

Sumei Fan[1,†] , Deyun Zhang[2,†], Yue Wang[2,†], Shijia Geng[2], Kun Lu[3], Meng Sang[2], Weilun Xu[2], Haixue Wang[4], Qinghao Zhao[5], Chuandong Cheng[6,7], Peng Wang[8], Shenda Hong[4,9,10,11,*]

1. College of Integrative Chinese and Western Medicine, Anhui University of Chinese Medicine, Hefei, China.
2. HeartVoice Medical Technology, Hefei, China.
3. Department of Electrocardiogram, The first Affiliated Hospital of Anhui Medical University, Hefei, China.
4. National Institute of Health Data Science, Peking University, Beijing, China.
5. Department of Cardiology, Peking University People's Hospital, Beijing, China.
6. Department of Neurosurgery, The First Affiliated Hospital of University of Science and Technology of China, Hefei, China.
7. Division of Life Sciences and Medicine, University of Science and Technology of China, Hefei, China.
8. Key Laboratory of Xinan Medicine of Ministry of Education, Anhui University of Chinese Medicine, Hefei，China.
9. Institute of Medical Technology, Peking University Health Science Center, Beijing, China.
10. Institute for Artificial Intelligence, Peking University, Beijing, China.
11. Department of Emergency Medicine, Peking University First Hospital, Beijing, China.

† Sumei Fan, Deyun Zhang and Yue Wang contributed equally to this article.

*Correspondence: Shenda Hong, hongshenda@pku.edu.cn

# Abstract

**Background:** Accurate and continuous electrocardiogram (ECG) parameter measurement outside hospital environments is essential for real-time cardiac health monitoring and telemedicine applications. On-device computation of single-lead ECG parameters enables timely assessment without reliance on centralized data processing, offering a pathway toward personalized, ubiquitous cardiac care. However, comprehensive validation of such computational methods across heterogeneous real-world populations remains limited.

**Methods:** We conducted a real-world validation using two datasets. HeartVoice-ECG-lite included 369 participants with single-lead ECG recordings annotated by two doctors. PTB-XL/PTB-XL+ comprised 21,354 patients with 12-lead ECG recordings and doctors' diagnostic annotations. FeatureDB (https://github.com/PKUDigitalHealth/FeatureDB), an on-device algorithm, was applied to compute PR, QT, and QTc intervals from single-lead signals. Accuracy was assessed using mean absolute error (MAE), correlation, and Bland-Altman analysis against doctor annotations. Diagnostic performances for first-degree atrioventricular block (AVBI, based on PR) and long QT syndrome (LQT, based on QTc) were benchmarked against commercial 12-lead systems (12SL, Uni-G) and an open-source algorithm (Deli), using AUC, accuracy, sensitivity, and specificity.

**Results:** FeatureDB-derived parameters showed high concordance with expert annotations, with MAEs comparable to inter-observer variability. Pearson correlations between FeatureDB and doctors ranged from 0.836 to 0.960 across parameters, closely matching inter-observer agreement. Bland-Altman analysis confirmed minimal bias and narrow limits of agreement. For AVBI detection, FeatureDB achieved AUC 0.787 (12SL: 0.859; Uni-G: 0.812; Deli:0.501). For LQT detection, AUC was 0.684 (12SL: 0.716; Uni-G:0.605; Deli:0.569), which performs comparably to commercial performance and is superior to existing open-source algorithms.

**Conclusion:** FeatureDB is capable of accurately and in real-time calculating various parameters of electrocardiograms from single-lead devices in the real-world out-of-hospital environment. This method achieves doctor-level accuracy in parameter estimation and reaches the level of commercial electrocardiogram machines in terms of abnormal heart detection. It supports scalable telemedicine applications, decentralized heart screening, and continuous monitoring in community and outpatient settings.

# Introduction

Cardiovascular diseases (CVDs) continue to pose a significant global health challenge, accounting for over 17.9 million deaths annually and representing nearly one-third of all global deaths[1]. Despite advances in diagnostics and therapy, the rising prevalence of CVDs—especially in aging populations and low- and middle-income countries—underscores the urgent need for accessible strategies enabling early detection and continuous monitoring[2]. As such, early identification and monitoring of cardiac abnormalities are critical to mitigating the societal and economic impact of these diseases. In recent years, single-lead devices have revolutionized cardiac monitoring by offering non invasive, continuous ECG parameters that bridges traditional intermittent assessments and patient care[3-5]. Their portability and real time data transmission enable the detection of transient arrhythmias and enhance patient engagement[4, 6]. These devices democratize cardiac care by expanding access to key ECG parameters — heart rate, PR interval, QRS duration, and QT interval — that are vital for diagnosing arrhythmias, ischemia, and other cardiovascular conditions, guiding risk stratification and therapy[7]. Precise parameters support large scale population health and personalized medicine initiatives[8]. To realize their clinical potential, however, the computational methods embedded in these devices must achieve accuracy comparable to doctor.

Recent studies have explored algorithmic approaches for calculating ECG parameters from wearable devices, leveraging signal processing and machine learning to improve precision under conditions of noise, motion, or low signal quality[9-11]. Ho et al.[9] demonstrated that Apple Watch and Garmin Forerunner provide heart rate estimates closely aligned with ECG parameters during exercise. Hwang et al.[10] confirmed the overall accuracy of consumer wearables in measuring heart rates during supraventricular tachycardia (SVT), though performance varied by device. Lu et al.[11] reported strong correlation ($r=0.84$, $p<0.01$) between single-lead and conventional ECG in heart rate parameters, suggesting potential reliability for arrhythmia detection. While these findings highlight progress in wearable cardiac monitoring[12,13], most prior studies have focused on limited parameters or small, controlled cohorts. Comprehensive validation of single-lead devices in real-world, heterogeneous populations remains lacking. In particular, there is a need to assess their performance in calculating multiple ECG parameters simultaneously, beyond simple heart rate estimation, and to benchmark these outputs against both ECG machines and expert evaluations.

To address this gap, this study aims to conduct a comprehensive real-world validation of ECG parameter calculation methods (Feature Database, FeatureDB) tailored for single-lead monitoring devices in real-wortld[14-16] (Figure 1). The primary aim is to assess the concordance and correlation between ECG parameters obtained from single-lead devices and observed by the doctor, leveraging a collected dataset from the real world. Additionally, the research systematically compares the performance of single-lead device-based parameter calculations against doctor assessments across multiple datasets, including a publicly available benchmark dataset and proprietary data from single-lead devices.This work aims to establish the accuracy, consistency, and clinical validity of single-lead device–based ECG parameter computation, providing evidence for their deployment in decentralized cardiac screening and long-term health management.

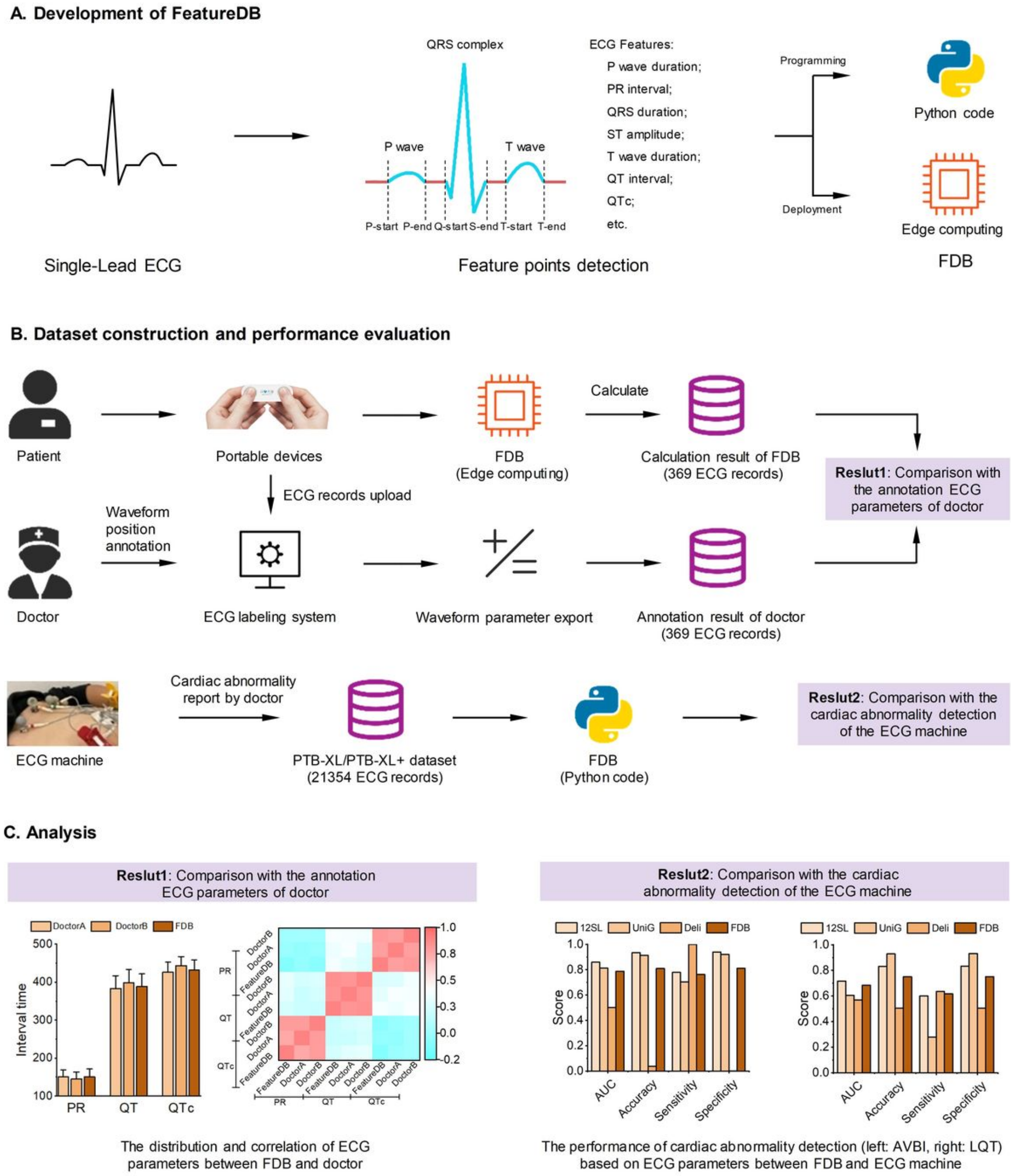

Figure 1. Overview of the study design. (A) The FDB was developed to extract ECG parameters from single-lead ECG signals and implemented as a scalable on-device analytical service. (B) The performance of the FDB was evaluated from two complementary perspectives: quantitative ECG parameters (benchmarking against doctors' annotations) and cardiac abnormality identification (benchmarking against standard 12-lead ECG machines). (C) The distribution and correlation of ECG parameters between the FDB outputs and doctors' assessments are illustrated. Additionally, the diagnostic performance of the FDB for detecting AVBU and LQT, based on derived ECG parameters, is compared with that of the ECG machine. FDB, FeatureDB; ECG, electrocardiogram; AVBI, first degree atrioventricular block; LQT, Long QT syndrome; PR, PR interval; QT, QT interval; QTc: corrected QT interval.

# Methods

## Study design

This study conducted a comprehensive real-world validation of ECG parameter calculation methods for single-lead monitoring devices, comparing their performance with both doctor and standard 12-lead ECG machines. Two distinct datasets were used (Table 1, Figure 2). The first single-lead dataset, HeartVoice-ECG-lite, comprised ECG recordings collected via WenXinWuYang[14-16] from 369 participants. This dataset included real-world ECG signals annotated with waveform parameters by two doctors. The second dataset (PTB-XL and PTB-XL+[17,18]) contained ECG parameters derived from two commercial ECG analysis software systems (12SL and Uni-G) and an open-source algorithm (Deli), along with doctor-reported cardiac abnormality labels. These datasets provided complementary perspectives: HeartVoice-ECG-lite highlighted the practical applications of emerging wearable technologies and enabled evaluation of discrepancies between FeatureDB-derived ECG parameters and expert annotations, whereas PTB-XL[17] and PTB-XL+[18] reflected real-world clinical practice and served as standardized benchmarks for assessing FeatureDB's capability in cardiac abnormality detection. By integrating these diverse data sources, this study robustly evaluated the performance of single-lead ECG parameter calculation methods under both controlled and real-world conditions, underscoring the potential of single-lead devices to enable scalable and accessible cardiac health monitoring. The study was approved by the Biomedical Ethics Committee of Peking University ( approval number: IRB00001052-23189). Given its real-word retrospective nature, the requirement for informed consent was waived.

Table 1. Patient characteristics. ECG: electrocardiogram.

| | HeartVoice-ECG-lite ($N$ = 369) | PTB-XL/PTB-XL+ ($N$ = 21354) |
|---|---|---|
| Age (year) | 42 (17) | 60 (17) |
| Female (%) | 182(49.3) | 10458 (47.9) |
| Height (cm) | 168 (9) | 167 (11) |
| Weight (kg) | 70 (15) | 71 (16) |

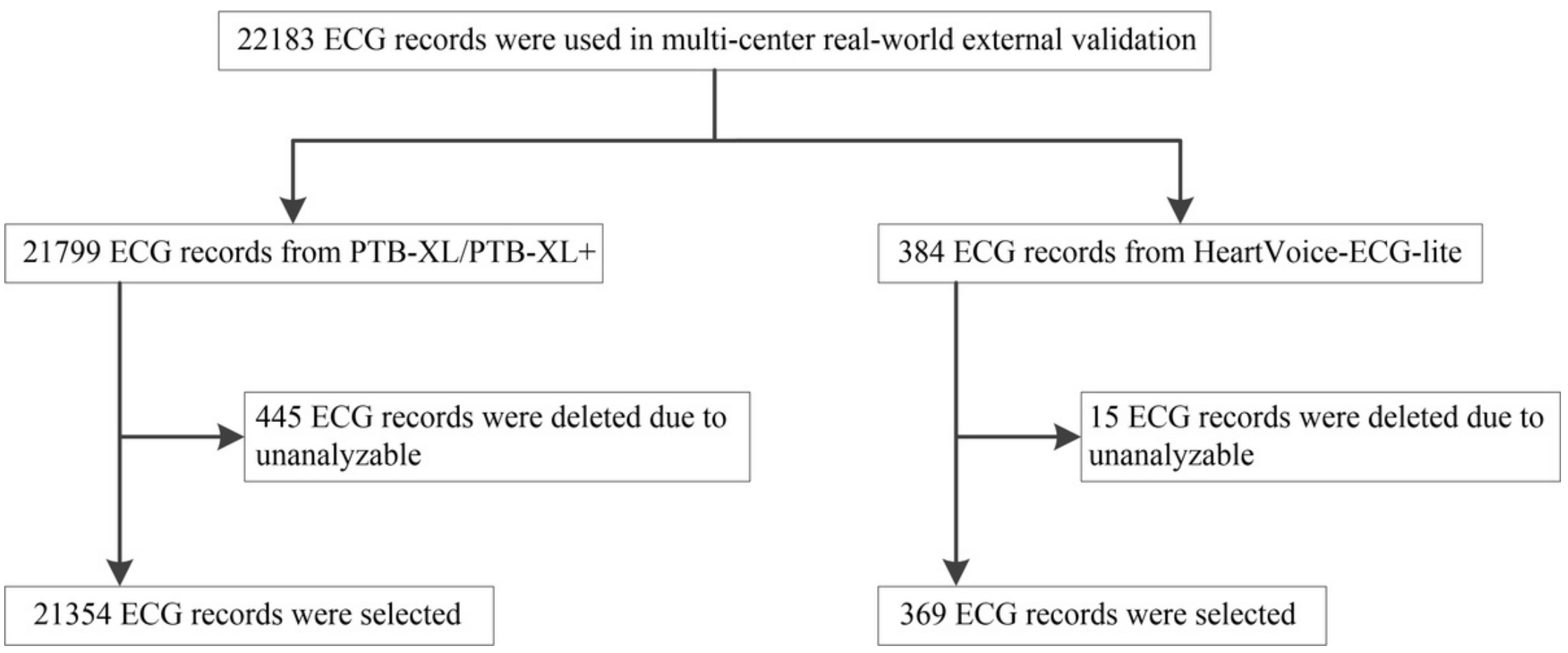

Figure 2. Patient flow diagram of this study. ECG: electrocardiogram.

## Feature Database

The FeatureDB (https://github.com/PKUDigitalHealth/FeatureDB) is a dedicated method developed for the computation of ECG parameters from single-lead ECG signals acquired by single-lead devices [14-16]. This method employs a multi-stage approach to accurately extract and calculate clinically relevant ECG parameters. Initially, FeatureDB detects the peak positions of the primary waveform components, namely the P wave, QRS complex, and T wave, by employing robust peak detection algorithms. This initial detection is critical for establishing a reliable foundation for subsequent analyses. Following the identification of the waveform peaks, the method applies multiple wavelet transforms to accurately determine the start and end positions of the P wave, QRS complex, and T wave. Wavelet transforms are particularly advantageous in this context due to their ability to analyze signals at various frequency scales, thus providing enhanced sensitivity to subtle changes in the ECG signal morphology. The refined delineation of waveform boundaries obtained through the wavelet-based approach allows for precise localization of key parameters, which is essential for the accurate computation of ECG parameters such as the PR interval, QRS duration, QT interval, and QTc. Subsequently, FeatureDB applies established calculation rules to the delineated waveform parameters to derive the desired ECG parameters. These calculation rules are based on clinically validated definitions and ensure that the computed parameters are consistent with standard diagnostic criteria. The integration of robust peak detection with multi-scale wavelet analysis enables FeatureDB to effectively handle the inherent noise and variability present in single-lead ECG recordings. Furthermore, the method is designed to be computationally efficient, making it well-suited for real-time or real-time applications in wearable technology.

## Single-lead device and doctor annotation

The HeartVoice-ECG-lite dataset was a single-lead ECG dataset recorded using the WenXinWuYang single-lead device in real-world cardiac health monitoring scenarios. This dataset underwent meticulous annotation by two doctors (Doctor A and Doctor B) using a specialized annotation system designed for precise identification of ECG waveform positions. The annotation process prioritized the accurate delineation of key waveform parameters, such as P waves, QRS complexes, and T waves, ensuring high reliability and clinical relevance of the annotations. Each doctor independently reviewed the data, and any discrepancies in annotations were resolved through consensus discussions to maintain consistency and minimize subjective bias. This rigorous approach aimed to provide a gold standard reference for evaluating the performance of ECG parameter calculation methods derived from single-lead devices. By incorporating expert-level annotations, the HeartVoice-ECG-lite dataset offers a valuable resource for benchmarking algorithm accuracy in diverse and practical settings, highlighting its importance in advancing the field of wearable technology for scalable and reliable cardiac health monitoring.

## PTB-XL and PTB-XL+ dataset

The PTB-XL[17] and PTB-XL+[18] dataset served as a cornerstone for our analysis by providing a comprehensive array of ECG parameters from 12-lead ECG machine and cardiac abnormality report by doctor. The PTB-XL and PTB-XL+ comprises a large collection of clinical 12-lead ECG recordings with rich diagnostic annotations derived from a diverse patient population. These recordings offer a robust benchmark for evaluating ECG analysis techniques. In parallel, two commercial ECG machine analysis software packages—12SL (GE Healthcare)[19] and Uni-G (The University of Glasgow ECG Analysis Program)[20]—were utilized; these platforms are widely adopted in clinical settings for their automated interpretation of ECG signals. Additionally, we incorporated Deli[21], an ECG analysis algorithm developed by a reputable German team, to further enhance the diversity and reliability of our comparative framework. The PTB-XL+ dataset extends the original PTB-XL by integrating computed ECG parameters, including detailed interval parameters and waveform parameters. This enriched dataset not only deepens the diagnostic information available but also enables rigorous external validations, ensuring that our single-lead ECG parameter calculation method is benchmarked against both established commercial solutions and expert doctor assessments.

## Outcomes

This study evaluated the accuracy of FeatureDB in calculating ECG parameters in real-world settings from two core perspectives. First, from the ECG parameters perspective, FeatureDB-derived results were compared with cardiologist annotations, and the distribution, mean absolute error (MAE), agreement, and correlation between the two were analyzed to assess parameters performance. Second, from the cardiac abnormality detection perspective, FeatureDB was compared with commercial 12-lead ECG machines, reporting the area under the receiver operating characteristic curve (AUC), accuracy, sensitivity, and specificity in detecting cardiac abnormalities, using first degree atrioventricular block (AVBI) and long QT syndrome (LQT) as representative examples.

## Statistical analysis

Continuous variables conforming to normal distribution are presented as mean ± standard deviation. Categorical variables were presented as numbers and percentages. Pearson correlation was used for continuous variables conforming to normal distribution. Bland-Altman plots were used to assess the level of consistency between two methods or devices, with mean and 1.96 standard deviation (SD). The diagnostic performance of FeatureDB were assessed by using (AUC, accuracy, sensitivity, and specificity. All data were analyzed by Origin 2018.

$$Accuracy = \frac{TP + TN}{TP + FP + TN + FN}$$

$$Sensitivity = \frac{TP}{TP + FN}$$

$$Specificity = \frac{TN}{FP + TN}$$

# Results

A total of 21,723 participants were included across the two datasets analyzed in this study (Table 1). The HeartVoice-ECG-lite cohort comprised 369 individuals (mean age 42 ± 17 years, 49.3% female), representing real-world users of single-lead wearable devices in community and ambulatory monitoring settings. In contrast, the PTB-XL/PTB-XL+ dataset included 21,354 patients (mean age 60 ± 17 years, 47.9% female) collected from clinical 12-lead ECG systems, serving as a reference for benchmarking FeatureDB against established diagnostic platforms. Anthropometric characteristics, including height and weight, were comparable between the two cohorts. Collectively, these datasets provided complementary perspectives—HeartVoice-ECG-lite emphasizing practical deployment in low-resource, real-world contexts, and PTB-XL/PTB-XL+ reflecting clinical diagnostic standards for robust external validation.

The ECG parameters calculated by the FeatureDB method demonstrated high concordance with the expert annotations provided by two doctors (Doctor A and Doctor B) in the HeartVoice-ECG-lite dataset. As illustrated in Figure 3A, the mean values and standard deviations for the PR interval, QT interval, and QTc interval were closely aligned across all three sources. Specifically, the mean PR interval estimations were nearly identical, reflecting robust detection of P-wave and QRS complex boundaries. While the QT and QTc intervals exhibited a slightly higher but clinically acceptable degree of variability, the overall consistency confirms that FeatureDB is capable of reliably delineating key waveform components and computing physiologically coherent ECG parameters from real-world, noisy single-lead ECG signals.

To quantify the accuracy of the algorithm, the MAE was calculated between FeatureDB-derived parameters and the cardiologists' annotations (Figure 3B). The MAE values remained consistently low across all parameters (PR, QT, and QTc) and were comparable to the inter-observer variability observed between Doctor A and Doctor B themselves. This quantitative evidence demonstrates that FeatureDB achieves doctor-level precision in ECG parameters computation. Figure 3C showcases a representative trace illustrating near-perfect agreement, where the FeatureDB-calculated PR, QT, and QTc parameters were virtually indistinguishable from the experts' markings. Conversely, Figure 3D highlights a challenging case where subtle changes in QRS-wave morphology resulted in minor discrepancies between the algorithm's calculation and the human annotations, though the difference remained clinically marginal. These combined findings support the robustness and high parameter fidelity of FeatureDB across diverse ECG signal qualities and morphologies, making it highly suitable for real-time monitoring applications in resource-constrained environments.

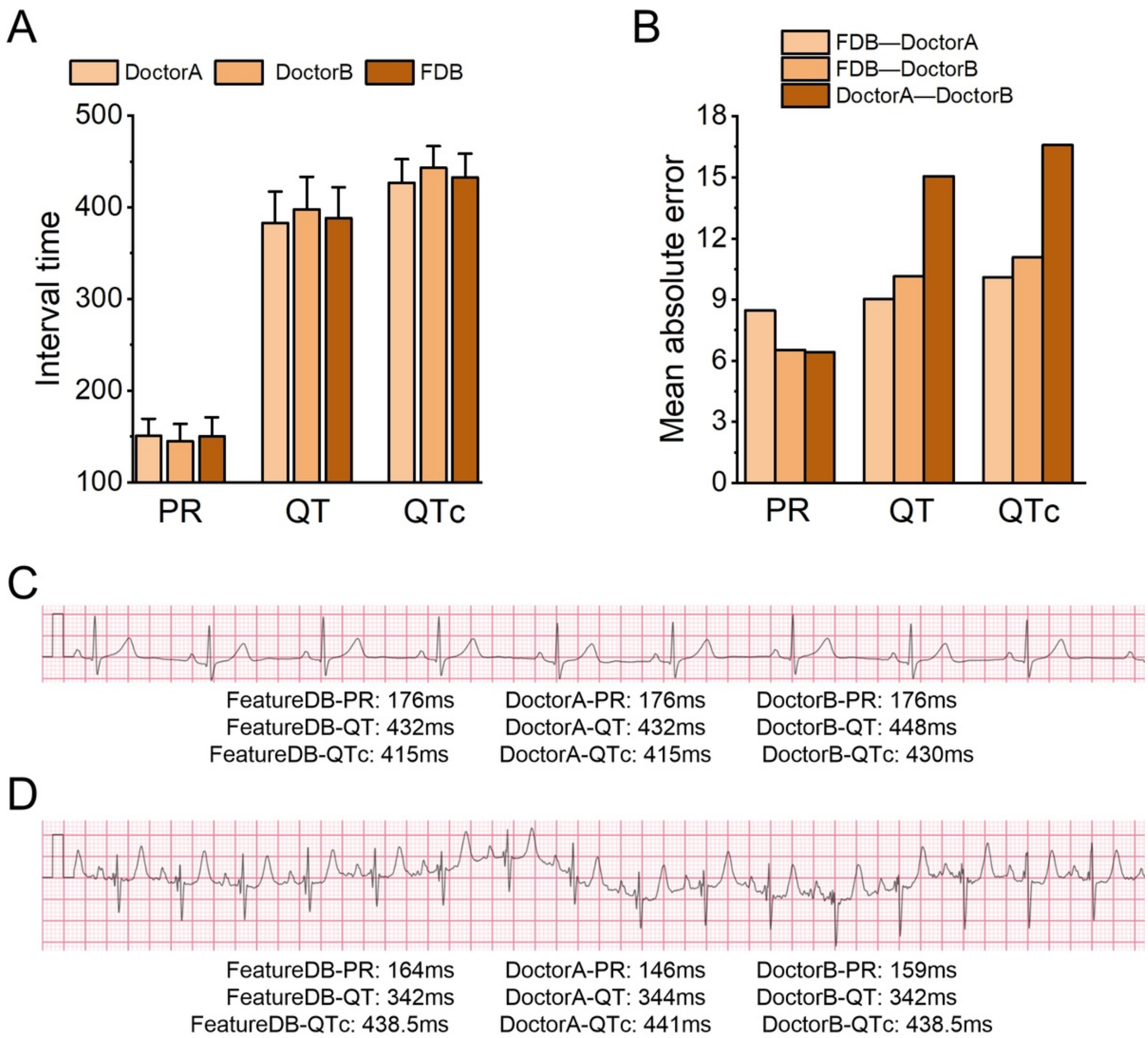

Figure 3. Comparison of ECG parameters measurements between FeatureDB and doctors. (A) Bar plots showing the mean PR, QT, and QTc measured by two doctors (Doctor A and Doctor B) and FDB. Error bars represent standard deviation. (B) MAE between measurements, comparing FDB with each doctor and inter-doctor variability. (C) and (D) Representative ECG trace from subject with stable rhythm and variable rhythm, showing measured PR, QT, and QTc intervals by FeatureDB and both doctors. ECG, electrocardiogram; FDB, FeatureDB; PR, PR interval; QT, QT interval; MAE: mean absolute error; QTc: corrected QT interval.

The correlation analysis further validated the precision of the FeatureDB algorithm, demonstrating a strong linear relationship between its output and the doctors' annotations across all tested ECG parameters (Figure 4). For the PR interval, FeatureDB showed a strong correlation with both Doctor A (r=0.836) and Doctor B (r=0.908) (Figures 4A and 4B). Notably, FeatureDB's correlation with Doctor B was nearly as high as the inter-observer correlation between the two doctors (r=0.911, Figure 4C). Similar high correlations were observed for the QT interval (FeatureDB vs. Doctor A: r=0.948; FeatureDB vs. Doctor B:

r=0.960; Figure 4D and 4E), which compared favorably to the inter-observer correlation (r=0.951, Figure 4F). Furthermore, the correlation for the QTc remained high (FeatureDB vs. Doctor A: r=0.892; FeatureDB vs. Doctor B: r=0.916; Figure 4G and 4H), again closely matching the inter-observer correlation (r=0.889, Figure 4I). The consistency across all parameters underscores the algorithm's capability to replicate expert-level parameters, a critical requirement for autonomous, on-device computation.

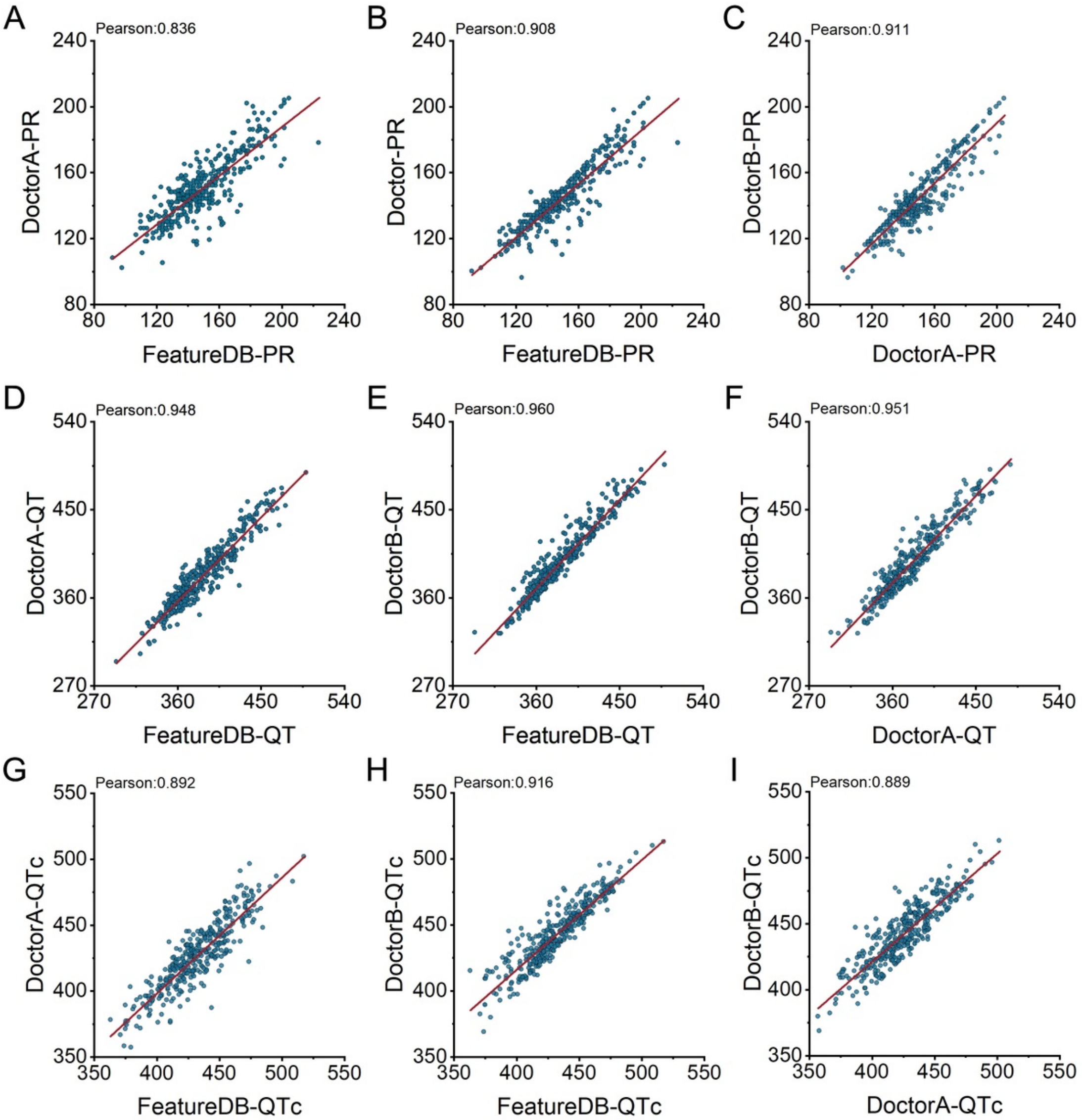

Figure 4. Correlation analysis of ECG parameters measurements between FeatureDB and doctors. (A) Correlation between FeatureDB-PR and DoctorA-PR. (B) Correlation between FeatureDB-PR and DoctorB-PR. (C) Inter-observer correlation between DoctorA-PR and Doctor B-PR. (D) Correlation between FeatureDB-QT and DoctorA-QT. (E) Correlation between FeatureDB-QT and DoctorB-QT. (F) Inter-observer correlation between DoctorA-QT and Doctor B-QT. (G) Correlation between FeatureDB-QTc and DoctorA-QTc. (H) Correlation between FeatureDB-QTc and DoctorB-QTc. (I) Inter-observer correlation between DoctorA-QTc and Doctor B-QTc. PR, PR interval; QT, QT interval; QTc: corrected QT interval.

Bland-Altman analysis was conducted to assess the agreement and potential systematic bias between FeatureDB and the expert annotations (Figure 5). For the PR interval (FeatureDB vs. Doctor A and Doctor B, Figure 5A and 5B), the mean differences (bias) were very close to zero, -0.715 ms and -5.076 ms, respectively, indicating minimal systematic error. The limits of agreement (LoA) for PR were tight, comparable to the LoA observed between Doctor A and Doctor B (Mean: -5.076 ms, Figure 5C). Analysis of the QT interval also showed strong agreement, though the bias for FeatureDB vs. Doctor A was slightly positive (5.098 ms, Figure 5D), while FeatureDB vs. Doctor B was negative (-9.767 ms, Figure 5E). Importantly, the spread of the data within the LoA for FeatureDB was generally tighter than the inter-observer LoA for the QT interval (Mean: -14.885 ms, Figure 5F), particularly when compared to Doctor B. Similar strong agreement with small bias was demonstrated for the QTc (FeatureDB vs. Doctor A: 5.730 ms; FeatureDB vs. Doctor B: -10.638 ms; Figure 5G and 5H). The minimal bias and narrow LoA across all parameters confirm that FeatureDB provides parameters that are interchangeable with expert annotations in clinical practice, further supporting its use in decentralized cardiac monitoring.

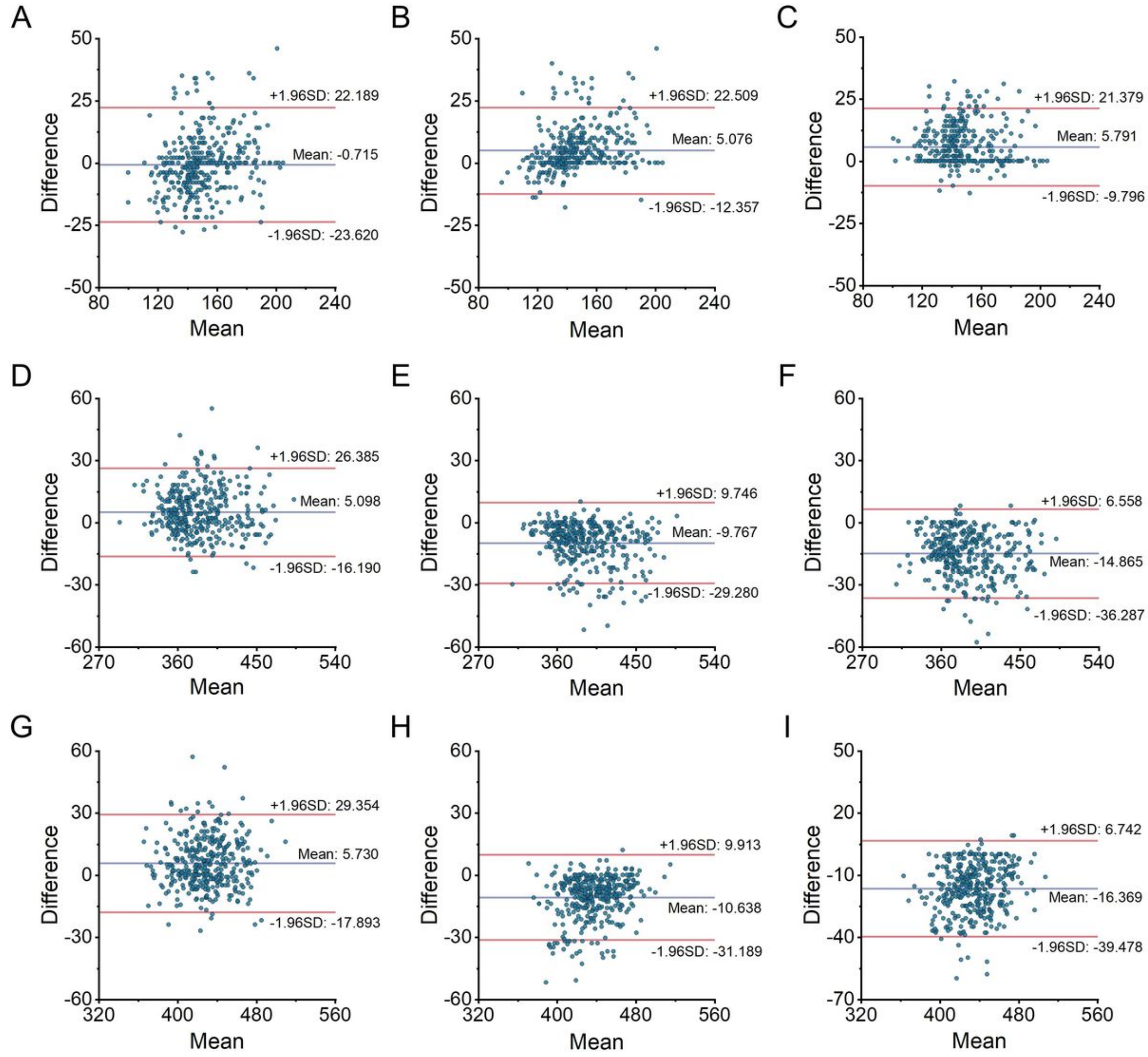

Figure 5. Bland-Altman analysis assessing the agreement of ECG interval measurements between FeatureDB and doctors.(A) Agreement between FeatureDB and DoctorA. (B) Agreement between FeatureDB and DoctorB. (C) Inter-observer agreement between DoctorA and DoctorB. (D) Agreement between FeatureDB and DoctorA. (E) Agreement between FeatureDB and DoctorB. (F) Inter-observer agreement between DoctorA and DoctorB. (G) Agreement between FeatureDB and DoctorA. (H) Agreement between FeatureDB and DoctorB. (I) Inter-observer agreement between DoctorA and DoctorB.The central purple line represents the mean difference, and the upper and lower red lines represent the 1.96 standard deviation. PR, PR interval; QT, QT interval; QTc: corrected QT interval.

We benchmarked the diagnostic performance of FeatureDB in identifying cardiac abnormalities against two commercial clinical ECG analysis systems (12SL and UniG) and an open-source algorithm (Deli) using the PTB-XL/PTB-XL+ dataset, focusing on AVBI (based on PR) and LQT (based on QTc) (Table 2). For AVBI detection, FeatureDB achieved a high AUC of 0.787 and an accuracy of 0.809, demonstrating strong capability. This performance was slightly lower than that of the two established 12-lead commercial systems (12SL AUC: 0.859, UniG AUC: 0.812) but significantly outperformed the Deli algorithm (AUC: 0.501). FeatureDB also maintained a clinically relevant sensitivity of 0.762 and specificity of 0.811 for AVBI. For LQT detection, FeatureDB yielded an AUC of 0.684 and an Accuracy of 0.75. While its AUC was lower than that of 12SL (0.716), it was superior to both UniG (0.605) and Deli (0.569). Crucially, the sensitivity of FeatureDB (0.617) for LQT was the highest than 12SL and Uni-G, indicating its strength in correctly identifying positive cases. These findings validate that FeatureDB, despite being designed for single-lead devices, provides clinically meaningful diagnostic performance comparable to, and in some metrics superior to, established multi-lead analysis platforms when applied to abnormality detection based on computed ECG parameters.

Table 2. Performance evaluation of FeatureDB, 12SL, Uni-G, and ECGDeli for Detecting LQT and AVBI in the PTB-XL and PTB-XL+ Datasets. AUC: the area under the receiver operating characteristic curve.

| | | 12SL | UniG | Deli | FDB |
|---|---|---|---|---|---|
| AVBI | AUC | 0.859 | 0.812 | 0.501 | 0.787 |
| | Accuracy | 0.934 | 0.913 | 0.037 | 0.809 |
| | Sensitivity | 0.779 | 0.703 | 0.999 | 0.762 |
| | Specificity | 0.939 | 0.921 | 0.001 | 0.811 |
| QTc | AUC | 0.716 | 0.605 | 0.569 | 0.684 |
| | Accuracy | 0.831 | 0.929 | 0.505 | 0.75 |
| | Sensitivity | 0.6 | 0.278 | 0.635 | 0.617 |

| Specificity | 0.832 | 0.932 | 0.504 | 0.751 |
|---|---|---|---|---|

# Discussion

This study addressed a critical gap in decentralized cardiac monitoring by providing a comprehensive, real-world validation of a computational method, FeatureDB, for accurately deriving key ECG parameters from single-lead devices. Our primary objective was to rigorously benchmark the performance of FeatureDB against both expert human annotations and established 12-lead clinical systems across diverse datasets, including the proprietary HeartVoice-ECG-lite collected in China and the public PTB-XL/PTB-XL+ dataset. The deployment of single-lead devices offers a vital, accessible pathway for early detection and continuous monitoring in resource-limited and community settings. Our results confirm that the ECG parameters (PR, QT, and QTc) calculated by FeatureDB demonstrate a high level of concordance and correlation with cardiologist assessments, with parameter accuracy comparable to inter-observer variability (Figures 3-5). Furthermore, when applied to cardiac abnormality detection, FeatureDB achieved clinically meaningful diagnostic performance for AVBI and LQT, rivaling or even surpassing some commercial and open-source algorithms (Table 2). These findings strongly validate the feasibility and reliability of on-device single-lead ECG parameter computation, enhancing the potential for these wearable technologies to revolutionize routine cardiovascular screening and management outside of traditional healthcare settings.

Accurate computation of ECG parameters enables early identification of conduction abnormalities (e.g., AVBI via PR) and repolarization abnormalities (e.g., LQT) in settings where standard 12-lead ECGs are unavailable or impractical. Previous work has shown that single-lead devices and wearables can reliably monitor heart rate and arrhythmia detection[4], but fewer studies have addressed multi-parameters (PR/QT/QTc) in real-world settings. For example, Hoek et al.[22] compared smartwatch QT intervals to standard ECGs, showing that QT and QTc parameters are feasible but less reliable than simple heart rate estimation. Our findings extend this by demonstrating robust performance across multiple parameters and heterogeneous recordings. From a therapeutic standpoint, continuous or frequent monitoring of QT or PR intervals could allow monitoring of drug-induced conduction changes or QT prolongation (e.g., in anti-arrhythmic, psychiatric, or oncology drugs) in outpatient and low-resource settings[23]. In resource-limited geographies, where access to cardiologists or full 12-lead ECG devices is constrained, deploying accurate single-lead ECG parameter computation offers a potential leap toward decentralized cardiac screening and monitoring. This not only supports risk stratification and early intervention but also empowers patient engagement and remote monitoring workflows[5]. The translation from ECG parameters accuracy to improved clinical outcomes remains to be established. Future work should quantify how algorithm-based ECG parameter monitoring impacts decision-making, referrals, therapy modification, and ultimately cardiovascular event reduction. However, our results provide a strong foundation for the clinical viability of algorithmic ECG parameters computation in wearable settings.

One of the key strengths of our study is the validation of FeatureDB across both a real-world, community single-lead dataset (HeartVoice-ECG-lite) and a large benchmark clinical 12-lead dataset (PTB-XL/PTB-XL+). This dual-dataset approach enhances generalizability–we show that the algorithm remains accurate under variable signal quality, device conditions, and real-world noise. Prior reviews have emphasized that many wearable ECG studies are conducted under controlled conditions or small cohorts[24], and that generalizability across populations, devices, and settings remains a major hurdle. Lim et al.[25] posits that real-world, long-term performance and diverse demographic inclusion are needed to move toward clinical adoption. Our usage of a Chinese single-lead ECG dataset and a large international dataset addresses this gap in part. Importantly, the low mean absolute errors and high agreement metrics in our single-lead cohort under ambulatory conditions suggest that device-algorithm performance is robust beyond laboratory settings. Moreover, the use of wearables in heterogeneous populations (e.g., variation in age, body habitus, signal noise) is crucial to ensure equitable utility in low- and middle-income countries (LMICs). The literature on wearable ECG in LMICs remains sparse. Thus, our work contributes to filling this evidence gap. Furthermore, our results may support regulatory and guideline discussions around algorithmic ECG parameters for non-traditional devices.

The advent of single-lead ECG monitoring paired with on-device or edge-embedded algorithms opens new possibilities for decentralized cardiovascular screening and continuous monitoring—especially relevant in low-resource settings. Traditional 12-lead ECG machines require infrastructure, trained personnel, and often hospital-based workflows; in contrast, wearable or patch-based single-lead systems enable ambulatory or home-based acquisition with minimal supervision[26]. Our findings indicating that doctor-level parameters accuracy support a paradigm in which primary care, community health workers, or even patients themselves can deploy monitoring, triggering referral or intervention when ECG parameter abnormalities (e.g., QTc or PR) are detected. Wearable device reviews have highlighted that single-lead ECG devices can bridge intermittent screening gaps and capture transient events[27]. The scalability is especially compelling: in aging populations or in LMICs with rising cardiovascular disease burden, such decentralized monitoring can reduce the burden on tertiary centers and enable earlier detection, thereby potentially reducing downstream morbidity and cost. From the public health viewpoint, large-scale data collection from wearables also offers population-level risk stratification and monitoring of intervention uptake.

However, successful implementation hinges on algorithmic trustworthiness, user adoption, device affordability, connectivity, data privacy, and health system integration. While our algorithmic results provide a strong technical basis, the next steps include implementation studies assessing workflow integration, user adherence, referral pathways, and cost-effectiveness in low-resource settings. Nevertheless, our work advances the vision of continuous "always-on" cardiac monitoring using single-lead devices and embedded analytics.

Despite the promising findings, our study has limitations that merit discussion and highlight future research directions. First, although our cohorts included a real-world single-lead dataset and a large benchmark dataset, certain populations were under-represented (e.g., pediatric subjects, extreme body habitus, severe arrhythmia burden). Other studies[3] show that in children, the accuracy of single-lead ECG parameters may be lower (Intra-class Correlation Coefficient for QTc 0.53). Second, although we validated the accuracy of ECG parameters and the detection of specific diagnoses (AVBI, LQT), the translation to long-term outcomes (e.g., arrhythmia incidence, adverse events) remains untested. Third, wearable/ambulatory monitoring introduces challenges of signal noise, motion artifact, adherence, and data management at scale[24].

This study demonstrates the feasibility and accuracy of on-device ECG parameter computation using FeatureDB in real-world, low-resource settings. The system achieved comparable performance to cardiologists in PR, QT, and QTc parameters and maintained robustness across varying signal qualities. These findings highlight its potential for scalable, real-time cardiac health monitoring and early disease detection outside conventional hospital environments.

## Acknowledgements

This study was supported by funds from the National Natural Science Foundation of China (No. 62102008, No. 62376256); the Joint Fund for Medical Artificial Intelligence (MAI2022Q011); the Fan Sumei scientific research start-up funds (DT2400000509).

## Data availability

Datasets included in this study are available from the corresponding author (Shenda Hong, hongshenda@pku.edu.cn) upon reasonable request.

## Code availability

Python code can be accessed in github: https://github.com/PKUDigitalHealth/FeatureDB.

## Declaration of competing interests

The authors declare that there are no competing interests.

## Author contributions

S.H. and D.Z. conceptualized this study. S.F., D.Z., and Y.W. led the data collection and reviewed the underlying data. S.F. led the data harmonization and statistical analysis. S.F., D.Z., Y.W., S.G., and S.H. wrote the first draft of the manuscript, which was substantially revised. All authors (S.F., D.Z., Y.W., S.G., K.L., M.S., W.X., H.W., Q.Z., C.C., P.W., and S.H.) made crucial contributions to several parts of the manuscript and had final responsibility for submission for publication.

# Reference

1. Kaptoge S, Pennells L, De Bacquer D, Cooney MT, Kavousi M, Stevens G, Riley LM, Savin S, Khan T, Altay S, Amouyel P. World Health Organization cardiovascular disease risk charts: revised models to estimate risk in 21 global regions. The Lancet global health. 2019 Oct 1;7(10):e1332-45.
2. Laranjo L, Lanas F, Sun MC, Chen DA, Hynes L, Imran TF, Kazi DS, Kengne AP, Komiyama M, Kuwabara M, Lim J. World Heart Federation roadmap for secondary prevention of cardiovascular disease: 2023 update. Global heart. 2024 Jan 22;19(1):8.
3. Ernstsson J, Svensson B, Liuba P, Weismann CG. Validation of smartwatch electrocardiogram intervals in children compared to standard 12 lead electrocardiograms. European Journal of Pediatrics. 2024 Sep;183(9):3915-23.
4. Hughes A, Shandhi MM, Master H, Dunn J, Brittain E. Wearable devices in cardiovascular medicine. Circulation research. 2023 Mar 3;132(5):652-70.
5. Sana F, Isselbacher EM, Singh JP, Heist EK, Pathik B, Armoundas AA. Wearable devices for ambulatory cardiac monitoring: JACC state-of-the-art review. Journal of the American College of Cardiology. 2020 Apr 7;75(13):1582-92.
6. Dhingra LS, Aminorroaya A, Oikonomou EK, Nargesi AA, Wilson FP, Krumholz HM, Khera R. Use of wearable devices in individuals with or at risk for cardiovascular disease in the US, 2019 to 2020. JAMA Network Open. 2023 Jun 1;6(6):e2316634-.
7. Khurshid S, Friedman S, Reeder C, Di Achille P, Diamant N, Singh P, Harrington LX, Wang X, Al-Alusi MA, Sarma G, Foulkes AS. ECG-based deep learning and clinical risk factors to predict atrial fibrillation. Circulation. 2022 Jan 11;145(2):122-33.
8. Bouzid Z, Al-Zaiti SS, Bond R, Sejdić E. Remote and wearable ECG devices with diagnostic abilities in adults: a state-of-the-science scoping review. Heart Rhythm. 2022 Jul 1;19(7):1192-201.
9. Ho WT, Yang YJ, Li TC. Accuracy of wrist-worn wearable devices for determining exercise intensity. Digital Health. 2022 Sep;8:20552076221124393.
10. Hwang J, Kim J, Choi KJ, Cho MS, Nam GB, Kim YH. Assessing accuracy of wrist-worn wearable devices in measurement of paroxysmal supraventricular tachycardia heart rate. Korean circulation journal. 2019 May 1;49(5):437-45.
11. Crossley GH, Boyle A, Vitense H, Chang Y, Mead RH, Connect Investigators. The CONNECT (Clinical Evaluation of Remote Notification to Reduce Time to Clinical Decision) trial: the value of wireless remote monitoring with automatic clinician alerts.

Journal of the American College of Cardiology. 2011 Mar 8;57(10):1181-9.

12. Rajakariar K, Koshy AN, Sajeev JK, Nair S, Roberts L, Teh AW. Accuracy of a smartwatch based single-lead electrocardiogram device in detection of atrial fibrillation. Heart. 2020 May 1;106(9):665-70.
13. Mannhart D, Lischer M, Knecht S, du Fay de Lavallaz J, Strebel I, Serban T, Vögeli D, Schaer B, Osswald S, Mueller C, Kühne M. Clinical validation of 5 direct-to-consumer wearable smart devices to detect atrial fibrillation: BASEL wearable study. Clinical Electrophysiology. 2023 Feb 1;9(2):232-42.
14. Hong S, Fu Z, Zhou R, Yu J, Li Y, Wang K, Cheng G. Cardiolearn: a cloud deep learning service for cardiac disease detection from electrocardiogram. InCompanion Proceedings of the Web Conference 2020 2020 Apr 20 (pp. 148-152).
15. Fu Z, Hong S, Zhang R, Du S. Artificial-intelligence-enhanced mobile system for cardiovascular health management. Sensors. 2021 Jan 24;21(3):773.
16. Li J, Aguirre AD, Junior VM, Jin J, Liu C, Zhong L, Sun C, Clifford G, Brandon Westover M, Hong S. An Electrocardiogram Foundation Model Built on over 10 Million Recordings. NEJM AI. 2025 Jun 26;2(7):AIoa2401033.
17. Wagner P, Strodthoff N, Bousseljot RD, Kreiseler D, Lunze FI, Samek W, Schaeffter T. PTB-XL, a large publicly available electrocardiography dataset. Scientific data. 2020 May 25;7(1):1-5.
18. Strodthoff N, Mehari T, Nagel C, Aston PJ, Sundar A, Graff C, Kanters JK, Haverkamp W, Dössel O, Loewe A, Bär M. PTB-XL+, a comprehensive electrocardiographic feature dataset. Scientific data. 2023 May 13;10(1):279.
19. Guide PS. Marquette® 12SL-ECG Analysis Program. 2005.
20. Macfarlane PW, Devine B, Clark E. The university of Glasgow (Uni-G) ECG analysis program. InComputers in Cardiology, 2005 2005 Sep 25 (pp. 451-454). IEEE.
21. Pilia N, Nagel C, Lenis G, Becker S, Dössel O, Loewe A. ECGdeli-An open source ECG delineation toolbox for MATLAB. SoftwareX. 2021 Jan 1;13:100639.
22. Hoek LJ, Brouwer JL, Voors AA, Maass AH. Smart devices to measure and monitor QT intervals. Frontiers in cardiovascular medicine. 2023 Nov 27;10:1172666.
23. Alam R, Aguirre A, Stultz CM. Detecting QT prolongation from a single-lead ECG with deep learning. PLOS Digital Health. 2024 Jun 25;3(6):e0000539.
24. Bayoumy K, Gaber M, Elshafeey A, Mhaimeed O, Dineen EH, Marvel FA, Martin SS, Muse ED, Turakhia MP, Tarakji KG, Elshazly MB. Smart wearable devices in cardiovascular care: where we are and how to move forward. Nature Reviews Cardiology.

2021 Aug;18(8):581-99.

25. Lim WH. Revolutionizing Healthcare: The Future of Wearable Single-Lead ECG Monitoring System. Korean circulation journal. 2024 Mar 1;54(3):154-5.
26. Zepeda-Echavarria A, van de Leur RR, van Sleuwen M, Hassink RJ, Wildbergh TX, Doevendans PA, Jaspers J, van Es R. Electrocardiogram devices for home use: technological and clinical scoping review. JMIR cardio. 2023 Jul 7;7:e44003.
27. Smith S, Maisrikrod S. Wearable Electrocardiogram Technology: Help or Hindrance to the Modern Doctor?. JMIR cardio. 2025 Feb 10;9(1):e62719.